\documentclass[final,3p,times,authoryear]{elsarticle}
\usepackage{amssymb}
\usepackage{amsmath}
\usepackage{graphicx}
\usepackage{xcolor}
\journal{Journal of Fluids and Structures}

\begin{document}

\begin{frontmatter}



%\title{Force partitioning and energy analysis applied to the analysis of shape effects in oscillating cylinders}

\title{Aeroelastic response of an airfoil to gusts: Prediction and control strategies from computed energy maps}


\author{Karthik Menon\corref{cor0}\fnref{label1}}
\cortext[cor0]{Corresponding author}
\ead{kmenon@jhu.edu}
\author{Rajat Mittal\fnref{label1}}
\ead{mittal@jhu.edu}

\address[label1]{Department of Mechanical Engineering, Johns Hopkins University, Baltimore, MD 21218, USA}

\begin{abstract}
A method to predict the aeroelastic pitch response of an airfoil to gusts is presented. The prediction is based on energy maps generated by high-fidelity fluid dynamic simulations of the airfoil with prescribed pitch oscillations. The energy maps quantify the exchange of energy between the pitching airfoil and the flow, and serve as manifolds over which the dynamical states of aeroelastic airfoil system grow, decay and attain stationary states. This method allows us to study the full nonlinear response of the system to large gusts, and predict the growth and saturation of aeroelastic pitch instabilities. We also show that the manifold topology in these maps can be used to make informed modifications to the system parameters in order to control the response to gusts. 
\end{abstract}


\end{frontmatter}


\section{Introduction}
\label{sec:intro}

The effect of gusts on the unsteady aerodynamics of wings has received significant attention due to its importance in aircraft design and control \citep{Anon2003GustLoads}. This is especially true of modern aircraft which employ lighter, flexible materials, as well as highly efficient large aspect-ratio wings \citep{Patil2004OnWings}. In addition, other gust-like disturbances, i.e. general external disturbances that induce structural perturbations, have importance in a variety of other fields of engineering. One example is the harvesting of energy from flow-induced flutter, where certain kinematic conditions have been shown to have higher energy harvesting efficiency than others \citep{Young2014,Xiao2014AFoils}. In these systems, external perturbations can be leveraged to drive the oscillations closer to optimal kinematic states, which can harvest energy from the flow more efficiently. In contrast to aerospace applications, these energy-harvesting systems operate with very large-amplitude oscillations \citep{Young2014,Xiao2014AFoils} where linear methods to analyze the response are no longer applicable. Hence, the large as well as small amplitude dynamical response of aeroelastic systems to perturbations has important practical applications. 

The bulk of previous work in this domain has focused on the transient forces and moments induced on a wing by an oncoming gust. In particular, the efficacy of analytical potential flow-based models, such as that of \cite{vonKarman1938AirfoilMotion} and its extensions, in predicting these experimentally measured force transients has been a topic of interest \citep{kuethe1939circulation,Granlund2014AirfoilFlows}, with mixed results especially in the presence of viscous effects and flow separation. This is driven by the need for simple and fast models of gust response that can be integrated with on-board control systems for aeroelastic applications. A particularly relevant model in this context is K{\"u}ssner's model for sharp-edge gusts, which, with the help of convolutional integrals has been explored as a potential model for the response to arbitrarily shaped gusts. \cite{Corkery2018OnRig} made one such comparison using experimental measurements of the response to a top-hat gust at $Re=40,000$, and showed that the K{\"u}ssner model predicted the peak lift well, but did not produce good predictions of the post-gust force recovery. Similarly, \cite{Perrotta2017UnsteadyGusts} compared experimentally measured forces induced by a transverse gust with predictions from thin-airfoil theory as well as a modified version of the K{\"u}ssner model. They found that neither of these models predicted the measured forces well. However, they showed that the peak force induced by the gust is very well predicted by the steady force at an effective angle of attack that includes the geometric angle of attack and the gust angle. Similar observations about the failure of the K{\"u}ssner model to predict the post-gust recovery were made by \cite{Biler2019ExperimentalEncounters}, who also showed that the effective angle of attack provides good predictions of the peak force. In fact, in our work, we show that the peak moments are also well predicted by the moments on a steady airfoil at an effective angle of attack given by the geometric angle plus the gust-induced angle, and we further use this in our analysis of the subsequent dynamic response of the airfoil. 

In addition to the experimental studies mentioned above, there have been computational efforts to analyze the gust response of wings using both, high-fidelity numerical simulations as well as reduced-order models (ROMs). As with the experimental studies, these efforts have primarily focused on analyzing the force response to gusts of various shapes and strength. This has led to the development of the field velocity method (FVM) to counter the effect of numerical dissipation due to upstream grid-stretching on the shape of the oncoming gust \citep{Singh1997DirectDynamics,Parameswaran1997IndicialCalculations}. However, due to the fact that this method only accounts for the effect of the gust on the wing, and not vice-versa, the split velocity method (SVM) was developed as an improvement to the FVM \citep{Wales2015PrescribedResponses}. Various comparisons of FVM and SVM have been made, and these methods have subsequently been used in a large number of computational studies to analyze the force response of the wing to incoming gusts \citep{Wales2015PrescribedResponses,Huntley2017AeroelasticFlows,Badrya2019ApplicationEncounter}. Furthermore, comparisons between these methods and K{\"u}ssner's model have yielded results similar to the comparisons with experiments \citep{Badrya2019ApplicationEncounter}.

While most of the existing literature on wing-gust interactions has focused on the force response, there has also been some work on investigating the aeroelastic response of wings to gusts. \cite{Huntley2017AeroelasticFlows} coupled SVM and FVM with a modal structural solver to study small amplitude deflections of an elastic wing, and compared results from the two methods. \cite{Yang2003NumericalAircraft} used a coupled solver to study small amplitude plunge and heave deflections of a wing due to harmonic as well as one-minus-cosine gusts. More recently, \cite{Doherty2019NonlinearExcitation} showed that the interaction of a gust with an aeroelastic cantilevered wing can in fact reduce the amplitude of limit-cycle oscillations observed, compared with the behaviour in the absence of the gust. 

A major focus in the coupled aeroelastic wing-gust domain has been on the development of simplified models to reduce the cost of such simulations. As mentioned earlier, this is again driven by the constraints of implementing on-board controllers for these systems. In this context, \cite{Tang2002LimitFlow} used the ONERA aerodynamic model by approximating the gust as a quasi-steady change in angle-of-attack, and coupled this with a nonlinear structural equation solver to study the deflection in a high aspect-ratio wing. \cite{Raveh2011Gust-responseRegime} developed a hybrid approach that involves computing responses to rigid sharp-edge gust in a high-fidelity simulation, computing the gust input forces due to arbitrary gust profiles via convolution, and then applying a linear aeroelastic feedback loop to compute the aeroelastic response. In a similar vein, various ROMs have been developed which aim to leverage the response calculations from a few expensive CFD simulations to develop simple models based on such convolutional methods, state-space approximations, etc. \citep{Gennaretti2004StudyWings,Zaide2006NumericalResponse,Raveh2007CFD-basedResponse,Wales2017Reduced-orderResponses}.

Due to the complexity and computational cost of studying the fully coupled dynamical response to arbitrary perturbations, the majority of the studies mentioned above employ simplifications and constraints, such as linearized force models or very small deflections. While these models have been shown to perform relatively well for aerospace applications that typically deal with very small deflections, they are not applicable to the analysis of the very large amplitude oscillations and nonlinear behaviour seen, for instance, in energy harvesting applications. During these large-amplitude oscillations, the effect of dynamic stall has been shown to be a major source of non-linearity in the system. Further, the accompanying growth and shedding of the leading-edge vortex causes large, transient aerodynamic forces, and poses significant challenges in terms of developing aeroelastic models that account for its effects \citep{Eldredge2019}. Consequently, the use of linearized aerodynamic models, and the focus on small-amplitude oscillations, ignores these non-linearities.

In this study, we present a method that uses data from fully coupled, high-fidelity numerical simulations to analyze and predict the large as well as small amplitude dynamical response of an aeroelastic system to gusts (or other arbitrary angle of attack perturbations). This makes the method presented here simultaneously applicable to aerospace and energy harvesting applications. Further, we also show with examples that this method allows us to make informed control decisions to alter the oscillation response of a system to perturbations, by simple manipulation of specific system parameters. In demonstrating this method, we focus exclusively on the dynamics of pitch-oscillations. This is to highlight its utility in accounting for aerodynamic non-linearities such as those associated with dynamic stall, as described above. In the domain of aeroelastic flutter, this single degree-of-freedom pitch oscillation is termed as ``stall flutter''. The characterization of stall flutter, and its associated non-linearities, has received significant attention in flow-induced pitching oscillations of airfoils \citep{dugundji1974stall,dugundji1975further,Poirel2008Self-sustainedNumbers,Dimitriadis2009BifurcationTunnel,Poirel2011ComputationalNumbers,Onoue2016VortexPlate,Menon2019} as well as two degree-of-freedom energy harvesting systems that are primarily driven by the pitching oscillations \citep{Zhu2009ModelingHarvester,Zhu2009ModeFoil,Young2013NumericalGeneration,Onoue2015LargePlate}. It must be noted that although the single degree-of-freedom model presented here is a simplification of classical/practical aeroelastic flutter applications, this method can also be used for certain higher degree-of-freedom applications, as will be described in section \ref{sec:energy_map}.

\subsection{Introduction to ``energy maps''}
\label{sec:energy_map}
\begin{figure}
  \centerline{\includegraphics[scale=1.0]{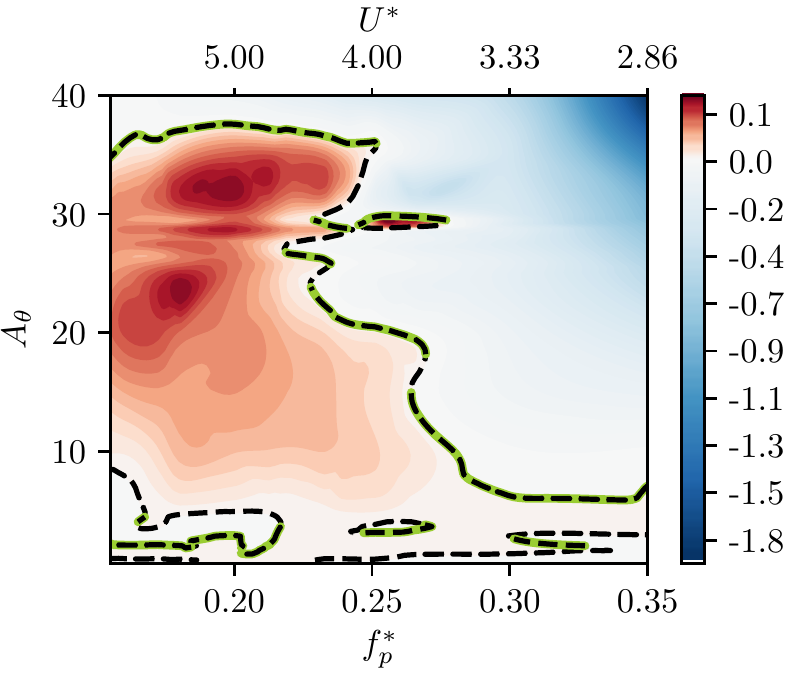}}
  \caption{Energy map reproduced from \cite{Menon2019} for an airfoil oscillating with mean angle $15^{\circ}$, about a point at $33\%$ of the chord, and at $Re=1000$ (for more details see ref. \citep{Menon2019}). The map shows contours of energy extraction (equation \ref{eq:energy_ext}) as a function of oscillation frequency and amplitude. The dashed curve shows the zero energy extraction contour, and the stable equilibria are highlighted in green.}
\label{fig:energy_hinge33}
\end{figure}
The method presented in this work is based on the ability to identify all possible limit cycle response branches (i.e. the "manifold") of an aeroelastic system, for a given range of kinematic and structural parameters \citep{Menon2019}. Once these response branches and their stability are known, we can use the instantaneous state of a system to predict and/or control the final stationary state of oscillation it will reach. Recent work has shown that the instantaneous amplitude growth/decay of an approximately sinusoidal flow-induced oscillating body (two-way coupled system) can be determined by calculating the energy extracted from the flow by the corresponding body under forced sinusoidal oscillations (one-way coupled) at matching kinematic conditions \citep{Morse2009,Bhat2013StallNumbers,Kumar2016Lock-inCylinder,Menon2019}. For the case of pitching airfoils, \cite{Menon2019} showed that forced sinusoidal pitch oscillations can be used to calculate the energy extracted by the airfoil from the freestream at a range of different amplitudes and frequencies of oscillation. For each of these cases the energy extraction coefficient ($C_E$) over an oscillation cycle is calculated as
\begin{equation}
  C_E = \int_{t}^{t+T} C_M \dot{\theta} dt
  \label{eq:energy_ext}
\end{equation}
where $\dot{\theta}$ is the angular velocity of the pitching airfoil, $T$ is the dimensionless period of oscillation, and $C_M$ is the coefficient of moment on the airfoil. This defines a map of energy extraction as a function of airfoil kinematics, which \cite{Menon2019} referred to as an ``energy map''. In figure \ref{fig:energy_hinge33} we show a sample energy map reproduced from this work \citep{Menon2019}, for an airfoil pitching about 33\% chord at a mean angle $15^{\circ}$ and $Re=1000$ (for more details see \cite{Menon2019}). Here $A_{\theta}$ and $f^*_p$ are the amplitude and frequency of pitch oscillation (defined in section \ref{sec:setup}), and each point on the map denotes the energy extraction coefficient for an airfoil pitching at that amplitude and frequency. On this map, kinematic states with positive (negative) energy extraction correspond to growing (decaying) oscillations, and (for zero structural damping) states with zero energy extraction correspond to equilibria. Furthermore, stable equilibria are given by states with $C_E=0$ and $dC_E/dA_{\theta}<0$ \citep{Morse2009,Menon2019}. Hence these stable equilibria define all possible stationary states for sinusoidally pitching airfoils which have structural frequencies in the frequency range of the energy map. In figure \ref{fig:energy_hinge33} we plot the equilibrium curve ($C_E=0$) using a dashed line, and stable equilibria ($C_E=0$ ; $dC_E/dA_{\theta}<0$) are highlighted in green along this line. The regions of nonzero energy extraction describe the non-equilibrium, transient behaviour of the flow-induced oscillating system and it was shown that this transient behaviour occurs at the natural frequency of the underlying oscillator. Hence, knowledge of the equilibrium and non-equilibrium states on these energy maps, together with the natural oscillation frequency of a system, allows us to predict the final stationary state response. 

We note that while the discussion here has focused on the energy map corresponding to a single degree-of-freedom system, this idea can, in principle, extended to higher degree-of-freedom systems. In such applications, the energy landscape would exist in a higher dimension, with each additional parameter corresponding to an added dimension in the energy landscape. For instance, in the purely pitching system described in this work, the energy extraction shown in figure \ref{fig:energy_hinge33} is a function of oscillation amplitude and frequency, i.e., $C_E(f^*_p,A_{\theta})$. On inclusion of heave oscillations with frequency $f^*_h$ and amplitude $A_h$, the energy landscape would be given by $C_E(f^*_p,A_{\theta},f^*_h,A_h)$. In these high-dimensional energy landscapes, the equilibria can still be identified as hypersurfaces of $C_E=0$ in the energy landscape, thus enabling analysis similar to what will be shown for the single degree-of-freedom case presented in this work.

In this work, we use the fact that the energy map allows us to identify all possible response branches of a system to predict the response of systems when they are pushed out of equilibrium by perturbations such as gusts. Using simple examples we also show that the structure of the energy map provides interesting and potentially useful insights into the response of the system to gusts.

\section{Computational method}
\subsection{Problem setup}\label{sec:setup}
\begin{figure}
  \centerline{\includegraphics[scale=1.0]{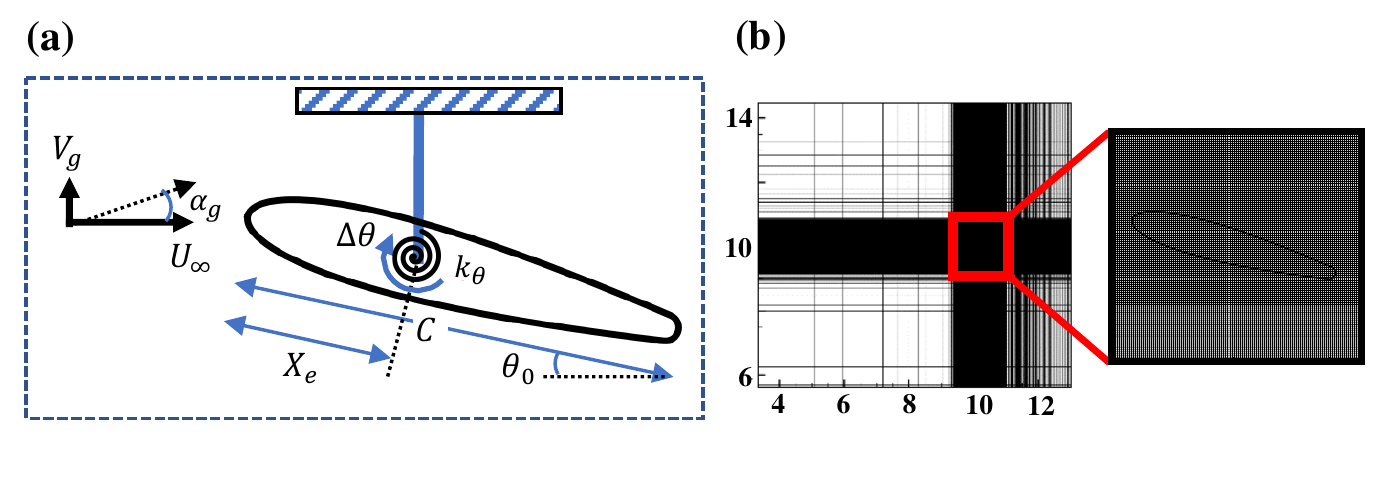}}
  \caption{(a) Schematic of the aeroelastic system used in this study; (b) Computational domain and close-up of the Cartesian computational grid around the airfoil. Note that the figure does not show the entire domain size, which is $21C \times 23C$.}
\label{fig:schematic}
\end{figure}
The two-dimensional flow-induced pitching oscillator studied in this work is a rigid NACA0015 airfoil, which is attached to an undamped linear torsional spring at a prescribed location along its chord. This simple elastic structural model is immersed in an incompressible fluid flow, and incoming gusts are modeled as vertical velocity perturbations of given magnitude and duration. A schematic of the computational model is shown in figure \ref{fig:schematic}(a). The airfoil used has a slightly rounded trailing edge to ensure the flow around it is well resolved, and we have verified that this has no significant effect on the aerodynamic forces. The governing equations for the fluid-flow as well as the elastic model are non-dimensionalized using the airfoil chord-length, $C$, as the characteristic length-scale, the freestream velocity, $U_\infty$, as the velocity scale and the convective freestream time-scale, $C/U_\infty$ as the characteristic time-scale. The fluid-flow is governed by the incompressible Navier-Stokes equations, the dimensionless form of which is as follows:
\begin{align}
  \nabla \cdot \vec{u} &= 0 \\
  \frac{\partial \vec{u}}{\partial t} + \vec{u}\cdot \vec{\nabla} \vec{u} &= - \vec{\nabla} p + \frac{1}{Re} \nabla^2 u
  \label{eq:navier-stokes}
\end{align}
Here $Re = \rho U_\infty C/\mu$ is the chord-based Reynolds number. The torsional elasticity of the pitching oscillator is governed by a forced spring-mass equation, with spring constant $k$, and moment of inertia $I$. As shown in figure \ref{fig:schematic}(a), this torsional spring is attached to the airfoil at a distance $X_e$ from the leading edge, which we refer to as the elastic axis. Furthermore, the equilibrium angular position of this torsional spring is referred to as $\theta_0$. The forcing to the system is provided by the airfoil's pitching moment, $M$. Here, the moment of inertia ($I$) and the pitching moment ($M$) are calculated about the elastic axis ($X_e$). The governing equation for the elastic model, non-dimensionalized using $C$ and $U_\infty$ as above, is hence
\begin{equation}
  I^* \ddot{\theta} + k^*(\theta-\theta_0) = C_M
  \label{eq:spring_eq}
\end{equation}
where $C_M = M/(0.5 \rho U^2_\infty C^2)$ is the coefficient of aerodynamic pitching moment about the elastic axis, $I^* = 2I/(\rho C^4)$ is the dimensionless moment of inertia about the elastic axis, and $k^* = 2k/(\rho U^2_\infty C^2)$ is the dimensionless spring stiffness. The dimensionless natural frequency of the elastic system is hence $f^*_n=\left({1}/{2\pi}\right) \sqrt[]{{k^*}/{I^*}}$, which is a parameter of interest in this study. It must be noted that in studies on aeroelastic flutter, this parameter is often written in terms of a reduced velocity $U^* = 1/f^*_n$. We report the location of the elastic axis in this work by its normalized value, $X_e^*=X_e/C$. Furthermore, the dimensionless frequency of pitch oscillation is denoted by $f^*_p=f_pC/U_{\infty}$, and the oscillation amplitude is denoted by $A_{\theta}$. 

\begin{figure}
  \centerline{\includegraphics[scale=1.0]{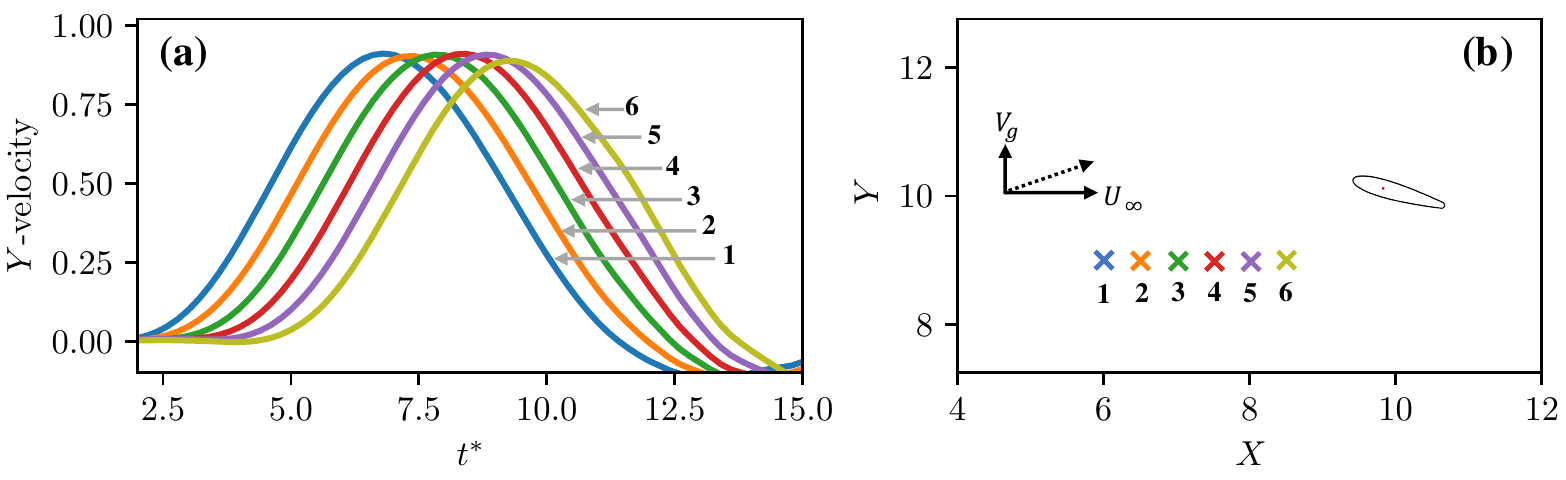}}
  \caption{(a) The profile of the incoming gust at 6 locations upstream of the airfoil, plotted as a time-series of vertical fluid velocity ($Y$-velocity) versus time; (b) Schematic showing the locations of the probe points at which $Y$-velocity is plotted in (a). The points are equally spaced from $X=6.0$ to $X=8.5$, with $Y=9.0$ for all points. The position of the airfoil is $X=10$, $Y=10$.}
\label{fig:gust_vel}
\end{figure}
The incoming gust is modeled using the so-called far-field boundary condition method (FBC). This is implemented as a time-varying vertical velocity of a prescribed strength and duration at the upstream boundary of the computational domain. In spite of limitations of this method  \citep{Singh1997DirectDynamics,Parameswaran1997IndicialCalculations,Wales2015PrescribedResponses}, we use this method here due to its simplicity and ease of implementation. Furthermore, as noted previously, the focus of this work is to demonstrate the use of energy maps \citep{Menon2019} in predicting the aeroelastic response to general gusts and gust-like perturbations and the simple gust model is sufficient for this purpose. 

In this work, the prescribed boundary condition at the upstream boundary takes the shape of a top-hat profile. As expected, diffusive effects upstream of the airfoil smoothen this imposed profile, hence the amplitude and duration of the top-hat profile are determined iteratively so as to obtain the required gust profile a short distance downstream of the boundary. The resulting gust profile upstream of the airfoil takes an approximately sine-squared (or one-minus-cosine) shape, which is quite commonly used in existing literature. The gust can therefore be approximated at a given stream-wise location as a travelling perturbation with a dimensionless velocity profile $v \approx V_g \sin^2(\pi t/T_g)$, which is defined for the time-interval $0<(t-T_0)<T_g$. Here $T_g$ is the dimensionless timescale of the gust, $V_g$ is the maximum vertical velocity in the gust, and $T_0$ is the time-instance at which the gust is incident at the particular stream-wise location. The spatio-temporal evolution this perturbation is shown in figure \ref{fig:gust_vel}(a) for a case with $V_g=0.9$, where the vertical velocity versus time at 6 upstream locations in the flow domain is plotted. The locations of the velocity probes in relation to the position of the airfoil is shown in figure \ref{fig:gust_vel}(b). Despite the effect of diffusion, we see that the gust profile is advected accurately for a considerable length in the regions upstream of the airfoil. Figure \ref{fig:gust_vel} shows this for a length $2.5C$ upstream of the airfoil ($X=6.0$ to $X=8.5$), but we have verified that this is true as far as $6C$ upstream of the leading edge of the airfoil. Hence the profile incident on the airfoil is practically unaffected by dissipation in this immediate upstream region, and we can report our results with respect to this profile. For the cases analyzed in this work, $T_g \approx 10$ (from fitting to a sine-squared profile), and the corresponding length of the gust, $L_g=T_gU_{\infty} \approx 10$.  The scale of the gust is hence much larger than the chord. The strength of the gust is reported in terms of an induced gust angle-of-attack, $\alpha_g = \arctan(V_g/U_{\infty})$, where $V_g$, the peak velocity in the gust profiles shown in figure \ref{fig:gust_vel}(a), is a parameter of interest. This is measured at the location $x=8$, $y=9$ where figure \ref{fig:gust_vel}(a) shows that the profile is well captured and upstream diffusive effects are no longer significant. This location is also chosen to ensure that the measured profile is not affected by the proximity to the airfoil.

\subsection{Numerical method}\label{sec:num_meth}
The coupled fluid-structure interaction problem is simulated using the sharp-interface immersed boundary method solver ``ViCar3D'' \citep{Mittal2008ABoundaries,Seo2011AOscillations}. With this method, sharp interfaces along the surface of the immersed body are preserved on a body-non-conformal Cartesian grid. This allows the simulation of large-amplitude motion of complex geometries using a simple non-adaptive Cartesian grid. This ability to simulate large-amplitude motions alleviates one major constraint in the previous investigations of flow-induced flutter cited earlier, which were mostly restricted by challenges associated with moving or deforming body-conformal grids. The fluid equations are solved using a second-order fractional-step method, and the pressure Poisson equation is solved using the geometric multigrid method. Second-order finite differences are used for all spatial derivatives, and the time-integration is performed using a second-order Adams-Bashforth method. Dirichlet velocity boundary conditions are enforced on the upstream boundary of the domain, and homogeneous Neumann conditions for velocity at all other boundaries. The pressure boundary conditions are homogeneous Neumann at all boundaries. This flow solver has been validated for a variety of stationary and moving boundary problems \citep{Ghias2007,Mittal2008ABoundaries,Seo2011AOscillations}, as well as for the particular set up of the problem in this study \citep{Menon2019}. 

The fluid-structure coupling in this work is implemented using explicit coupling, wherein the fluid flow and elastic equations are solved sequentially. The aerodynamic forces/moments are calculated on the Lagrangian marker points along the airfoil surface, and then passed to the equation governing the oscillations of the airfoil, i.e. equation \ref{eq:spring_eq}. We then use equation \ref{eq:spring_eq} to calculate the angular velocity of points along the airfoil surface, and use this to update their positions. The computational domain used in this work spans $21C \times 23C$, and the airfoil is placed $10C$ from the downstream boundary. A grid size of $384\times320$ points is used for all simulations in this work, and the resolution around the airfoil corresponds to about $125$ points along the chord. A zoom-in of the grid around the airfoil is shown in figure \ref{fig:schematic}(b). A grid convergence study is provided in \cite{Menon2019}, where the variation in the pitch deflection as well as the mean and RMS lift and moment coefficients were shown to be small for grids with $2.3\times$ and $3.6\times$ more grid points.

For all the cases discussed in this study, simulations are initialized with the airfoil at its equilibrium angle, $\theta_0$, and zero angular velocity. Further, the equilibrium angle is fixed to a value of $\theta_0=15^{\circ}$ throughout this work. The Reynolds number is fixed at $Re=1000$ for all cases discussed here. The moment of inertia is also fixed, at $I^*= 0.073$ when $\theta$ is measured in degrees. This choice of the moment-of-inertia about the elastic axis corresponds to a solid-to-fluid density ratio of $\approx 120$. The freestream velocity, $U_\infty$, is prescribed at the upstream boundary of the domain, and a vertical velocity is added to this upstream boundary condition for a fixed duration. This corresponds to a gust of length $L_g \approx 10$, as shown in figure \ref{fig:gust_vel}, and angle $\alpha_g$. For simplicity, this gust perturbation is prescribed early enough in the simulation to ensure that the amplitude of the airfoil's oscillation is small enough and therefore, its phase at the instant of the gust encounter can be ignored. Subsequently, the dynamics are allowed to evolve until a stationary state is attained. 

\section{Results}

In this section we discuss the gust-response of the system described above. As mentioned earlier, the focus is on the use of energy maps in predicting, and possibly enabling control of the response of an aeroelastically pitching airfoil to gusts. We illustrate this using various examples of airfoil-gust encounters, each highlighting different aspects of the utility of energy maps in such analyses.

\subsection{Predicting gust-induced aeroelastic flutter}
\label{sec:f0.19}
\begin{figure}
  \centerline{\includegraphics[scale=0.75]{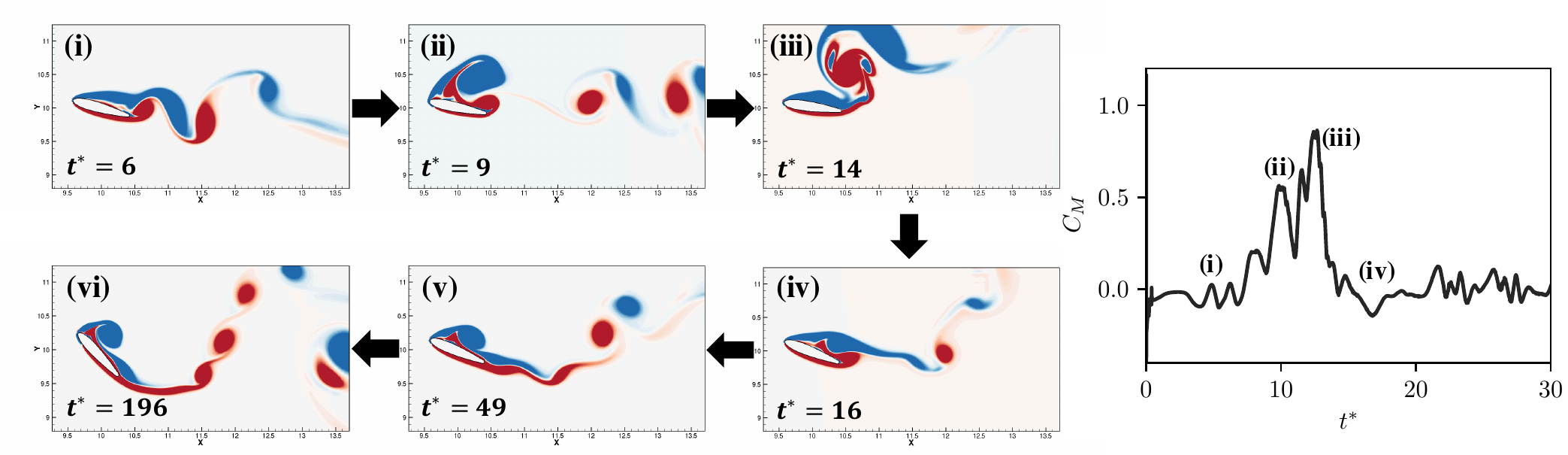}}
  \caption{For an airfoil encountering a gust of strength $\alpha_g=42^{\circ}$, and oscillating about $X^*_e=0.33$ with frequency $f^*_n=0.19$, (i)-(vi) show snapshots of the flow-field using contours of vorticity, at various time instances during and after the airfoil-gust encounter. Right pane shows the time-series of moment coefficient, with time-instances corresponding to snapshots marked as (i)-(iv). The corresponding time-series of pitch angle for this case is shown in figures \ref{fig:u40_timeseries}(b) and \ref{fig:u40_pitch_zoom}(b), with time-instances corresponding to the snapshots marked as (i)-(vi).}
\label{fig:snapshots}
\end{figure}

\begin{figure}
  \centerline{\includegraphics[scale=0.9]{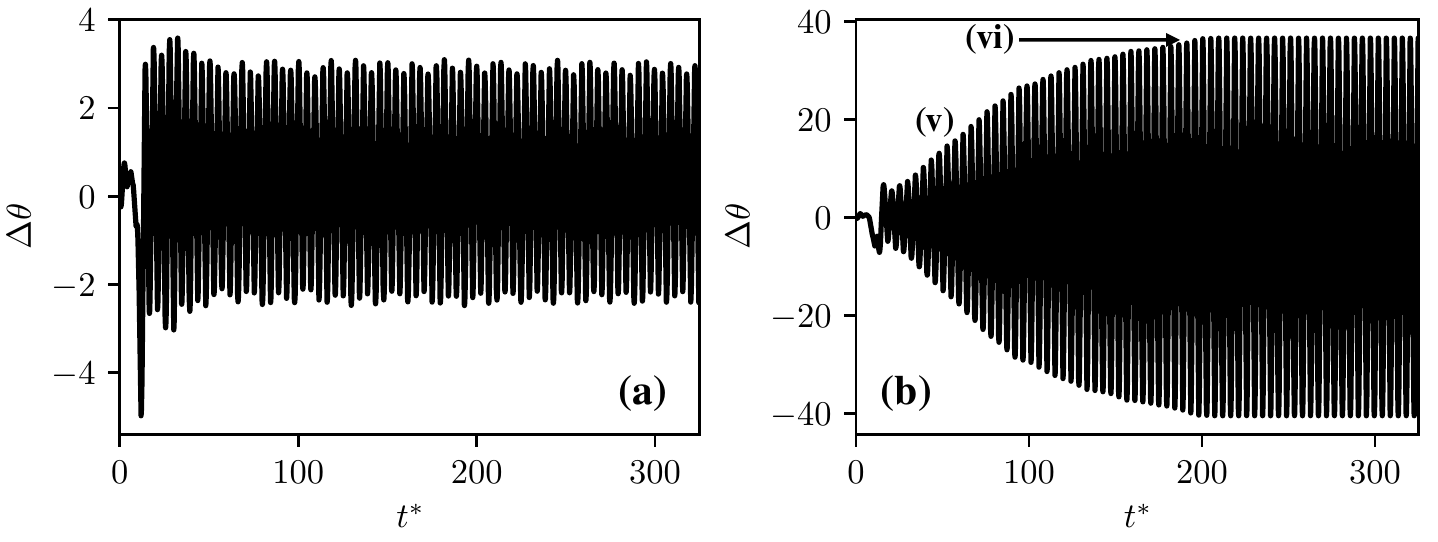}}
  \caption{Time-series plots of pitch angle for two airfoils with $f^*_n=0.19$ and $X^*_e=0.33$, encountering gusts of different strengths. (a) $\alpha_g=24.5^{\circ}$; (b) $\alpha_g=42^{\circ}$. The time-instances corresponding to snapshots in figures \ref{fig:snapshots}(v) and \ref{fig:snapshots}(vi) are shown in (b).}
\label{fig:u40_timeseries}
\end{figure}

\begin{figure}
  \centerline{\includegraphics[scale=0.9]{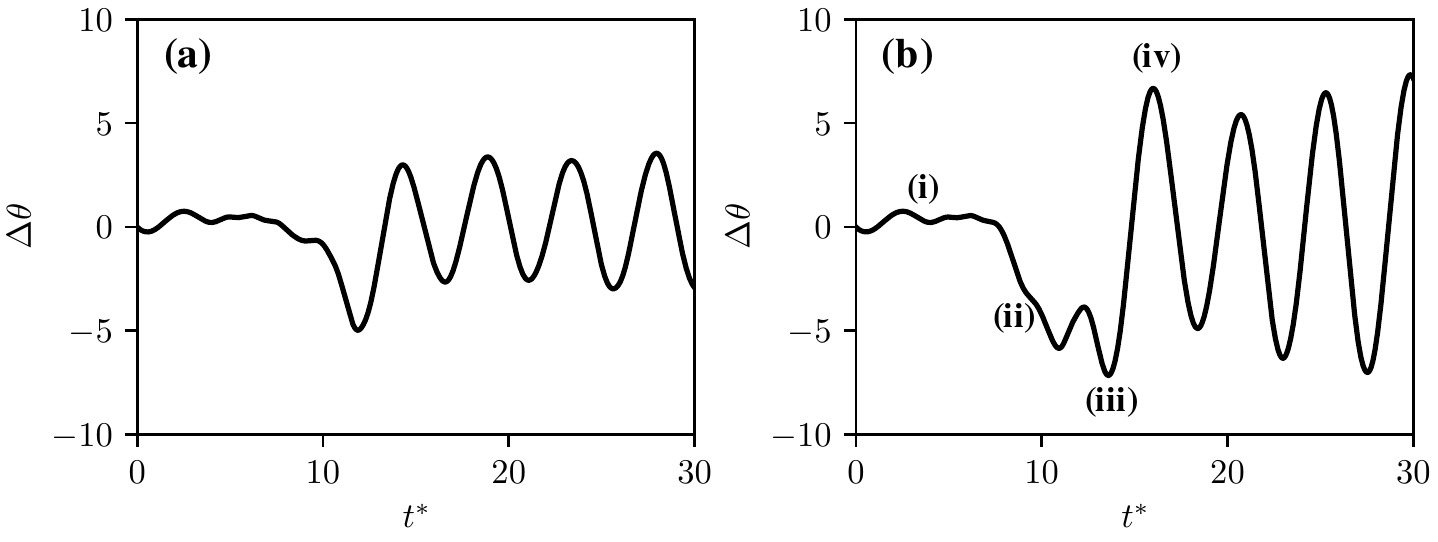}}
  \caption{(a) Zoom-in of the time-series plots in figure \ref{fig:u40_timeseries}, focusing on the initial pitch-deflection induced by the gust. The parameters for these cases are the same as in figure \ref{fig:u40_timeseries}, i.e., $f^*_n=0.19$, $X^*_e=0.33$, and gust strengths given by (a) $\alpha_g=24.5^{\circ}$; (b) $\alpha_g=42^{\circ}$. The time-instances corresponding to snapshots in figures \ref{fig:snapshots}(i)-(iv) are shown in (b). }
\label{fig:u40_pitch_zoom}
\end{figure}

We begin with a discussion of the flutter response of two airfoils with the same structural (i.e. elastic) properties, but encountering gusts of different magnitudes. The natural frequency of the elastic systems under consideration is fixed at $f^*_s=0.19$, which corresponds to a $U*=5.26$, and the elastic axis is at $X^*_e=0.33$. The incoming gust angles are $\alpha_g=24.5^{\circ}$ and $\alpha_g=42^{\circ}$. In figure \ref{fig:snapshots} we show snapshots of the flow at a few time instances during and after the wing-gust interaction for one of these cases, with $\alpha_g=42^{\circ}$. The flow in these snapshots is represented by contours of $Z$-vorticity (normal to the plane of the paper). On the right-pane of figure \ref{fig:snapshots} we also show the coefficient of pitching moment experienced by the airfoil during the gust encounter for this case. Further, the pitch deflection corresponding to the time instance at each snapshots is shown in figures \ref{fig:u40_timeseries}(b) and \ref{fig:u40_pitch_zoom}(b), which are plots of the pitch deflection time-series for this case. Figure \ref{fig:snapshots}(i) shows a snapshot of the flow before the gust interaction, where the airfoil is at angle $\theta \approx 15^{\circ}$, and we see the development of a von Karman vortex street in the wake. On initially encountering the gust, the airfoil experiences large-scale flow separation on the suction side, as seen in figure \ref{fig:snapshots}(ii). This is accompanied by the generation of a large moment transient, seen at time-instance (ii) in the plot of pitching moment, which induces an initial pitch deflection that drives the ensuing dynamics. We show snapshots of this initial deflection at the maximum pitch-down position in figure \ref{fig:snapshots}(iii) and at the maximum pitch-up position in figure \ref{fig:snapshots}(iv). The corresponding pitch deflections can be seen in figure \ref{fig:u40_pitch_zoom}(b), at the time-instances marked as (iii) and (iv), and is measured to be $\Delta \theta \approx 6^{\circ}$. The gust then advects downstream of the airfoil as the pitching dynamics continue to evolve. In the case being discussed here, the initial pitch deflection leads to further growth in the oscillation amplitude, which is shown in figures \ref{fig:snapshots}(v) and \ref{fig:snapshots}(vi). We see from figure \ref{fig:u40_timeseries}(b) that the amplitude of pitch oscillation is $\Delta \theta \approx 15^{\circ}$ during the oscillation cycle shown at time-instance (v). The oscillations are eventually seen to approach a limit-cycle, as shown at time instance (vi) in figure \ref{fig:u40_timeseries}(b), the amplitude of which is $\Delta \theta \approx 35^{\circ}$. Hence we see that even this simplified aeroelastic model is capable of showing large-amplitude limit-cycle oscillations. Further, the flow physics is influenced by large regions of separated flow, which precludes the use of linear aerodynamic models at these low Reynolds numbers.

Figure \ref{fig:u40_timeseries} shows time-series plots of the pitch deflection, $\Delta \theta$, for these cases with $f^*_n=0.19$ and gust strengths $\alpha_g=24.5^{\circ}$ and $\alpha_g=42^{\circ}$. We see for the case of $\alpha_g=24.5^{\circ}$ that the gust induces an initial pitch-up deflection of $\Delta \theta \approx 4^{\circ}$, after which the airfoil settles into limit cycle oscillations of amplitude $\Delta \theta \approx 3^{\circ}$. For the case with $\alpha_g=42^{\circ}$, which represents a gust that is almost $2\times$ stronger, we see a very different behaviour. After the initial deflection of $\Delta \theta \approx 7^{\circ}$, the amplitude of oscillation grows very rapidly to a stationary state of $\Delta \theta \approx 37^{\circ}$. Hence gusts that differ by a factor of $2\times$ in strength cause stationary state oscillation amplitudes that differ by $>10\times$. This highlights the inherent non-linearity in the system, despite the linearity of the underlying dynamical model. 

\begin{figure}
  \centerline{\includegraphics[scale=0.9]{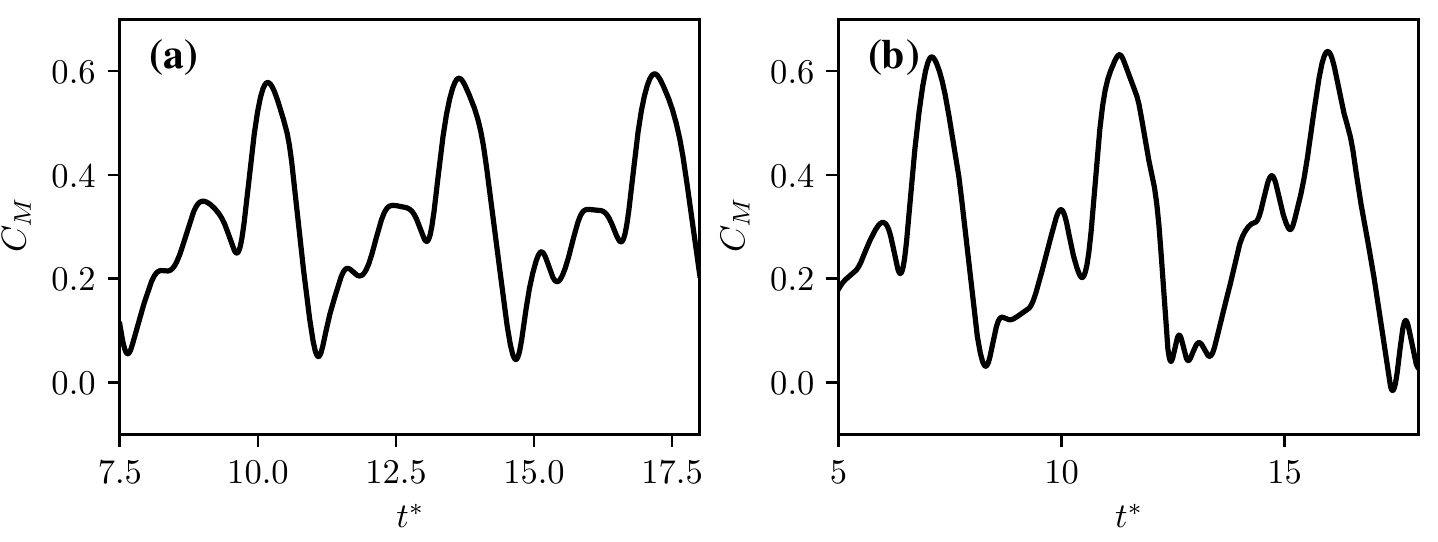}}
  \caption{Coefficient of moment about $X^*_e=0.33$, on airfoils at two static angles of attack given by (a) $\theta_s=39.5^{\circ}$; (b) $\theta_s=57^{\circ}$. Here $\theta_s = \theta_0+\alpha_g$ is the equivalent quasi-steady angle experienced during the initial gust encounter, for an airfoil at equilibrium angle $\theta_0$ and encountering a gust of strength $\alpha_g$.}
\label{fig:u40_moment_zoom}
\end{figure}
In spite of the order-of-magnitude difference in the stationary state oscillation amplitudes for these cases, the initial pitch deflection induced by the gust shows much less variation. Figure \ref{fig:u40_pitch_zoom} shows the time-series of pitch angle plots for the cases with $\alpha_g=24.5^{\circ}$ and $\alpha_g=42^{\circ}$, focusing on the initial deflection for each case. We measure the peak initial deflection in each of these cases to be $\Delta \theta = 4.98^{\circ}$ and $\Delta \theta = 5.85^{\circ}$ for $\alpha_g=24.5^{\circ}$ and $\alpha_g=42^{\circ}$ respectively. This suggests that a different, and possibly simpler, mechanism governs this initial deflection. Since the length-scale of the gust is much larger than the chord of the airfoil, it is reasonable to assume that the initial force and moment transient due to the gust can be approximated in a quasi-steady manner. Hence we expect that the initial pitch deflection should be governed by the pitching moment generated on an airfoil at a steady apparent angle of attack given by $\theta_s = \theta_0+\alpha_g$. For two cases considered here, with $\alpha_g=24.5^{\circ}$ and $\alpha_g=42^{\circ}$, the resulting steady angles of attack are $\theta_s=15^{\circ}+24.5^{\circ} = 39.5^{\circ}$ and $\theta_s=15^{\circ}+42^{\circ} = 57^{\circ}$ respectively. In figure \ref{fig:u40_moment_zoom} we show $C_M$ time-series plots at these steady angles $\theta_s$, for the same airfoil used in this study. Here $C_M$ is calculated about a location at $33\%$ of the chord. From this data, we can extract the peak $C_M$ at each $\theta_s$, which we refer to as $C_M'$. The value of $C_M'$ at $\theta_s=39.5^{\circ}$ and $\theta_s=57^{\circ}$ are $C_M' = 0.57$ and $C_M' = 0.63$ respectively. For airfoils with elastic stiffness $k^*$, the initial pitch deflection induced by the gust is expected to be given by $\Delta \theta'=C_M'/k^*$. As mentioned earlier, both these cases have the same elastic properties, with $k^* = 0.1054$. Hence $\Delta \theta'$, based on the quasi-steady assumption, comes out to be $\Delta \theta' = 5.41^{\circ}$ and $\Delta \theta' = 5.95$ respectively. We note that this is close to the observed values of initial deflection, $\Delta \theta = 4.98^{\circ}$ and $\Delta \theta = 5.85^{\circ}$, which confirms the validity of this quasi-steady assumption. As mentioned in the Introduction, a similar finding was made by \cite{Perrotta2017UnsteadyGusts} who showed via experiments that the initial lift-force transient was given by the force on a steady wing at an angle of attack given by $\theta_s$. 

We now move on to analyzing the final stationary state of the system that results from the interaction with the gust. In this work, as in \cite{Menon2019}, the current state is determined by the amplitude and frequency of the current oscillation cycle and the initial state is determined by the initial deflection induced by the gust. The energy map shows the energy extraction coefficient for the airfoil, $C_E$, at each state. When the system is at a state with $C_E \neq 0$, the amplitude of oscillation is expected to amplify or decay (as per the sign of $C_E$) until the system reaches a stable equilibrium. Based on this, and given the initial deflection induced by a gust (or a gust-like perturbation), we expect to be able to use the energy map to predict the final stationary state of the system. We first verify these ideas in the context of the gust response by comparing the observed response with the energy map-based predictions. The energy map used here is essentially the same as that presented in \cite{Menon2019}, except that the resolution of this plot has been improved by adding more simulation data.

\begin{figure}
  \centerline{\includegraphics[scale=1.0]{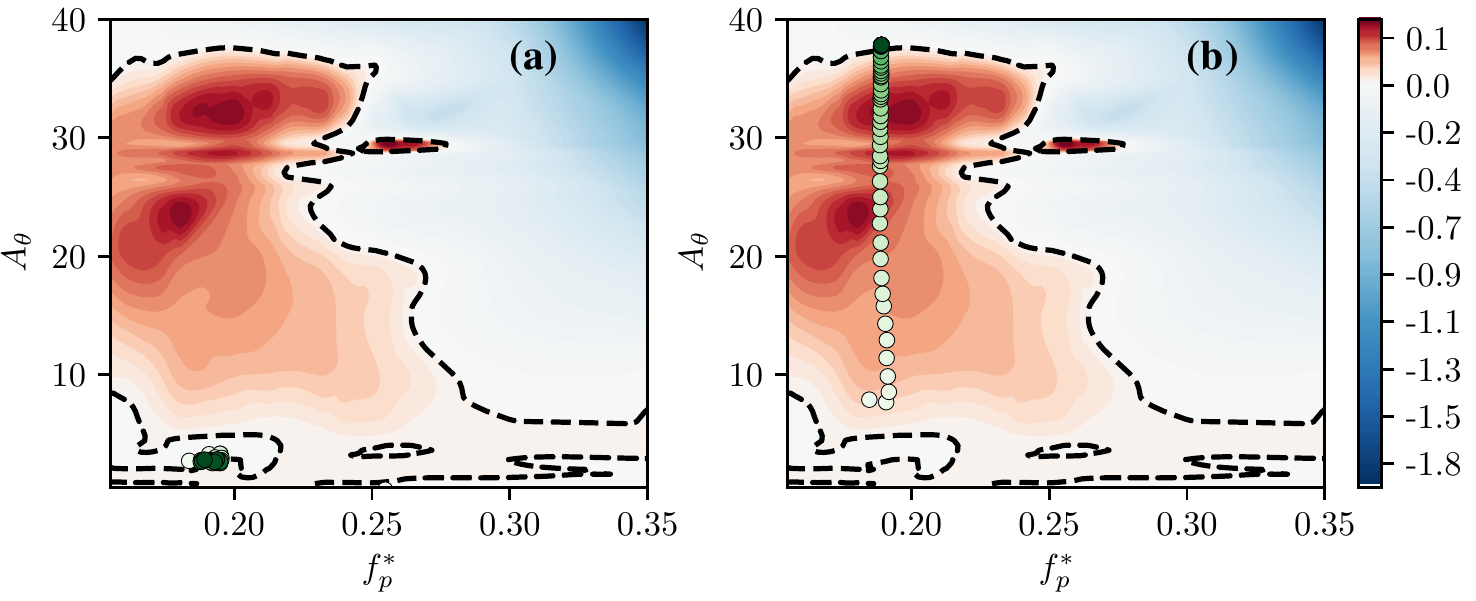}}
  \caption{(a) Frequency-amplitude trajectories for two cases with different gust strengths, plotted on the energy map for $X^*_e=0.33$. The trajectories, shown using circles, represent the evolution of the pitching dynamics for each case. The circles represent the cycle-averaged frequency and amplitude at every oscillation cycle. The darkness of the circles represent time, i.e., the darker circles represent the dynamics at later times in the system's evolution. The structural properties of the airfoils in these cases are $f^*_n=0.19$ and $X^*_e=0.33$. The gust strengths are (a) $\alpha_g=24.5^{\circ}$; (b) $\alpha_g=42^{\circ}$.}
\label{fig:u40_energy}
\end{figure}
In figure \ref{fig:u40_energy} we plot the energy map for an airfoil pitching about $X^*_e=0.33$, with equilibrium angle of attack $\theta_0=15^{\circ}$, and at $Re=1000$. Overlaid on the energy map, is the cycle-averaged oscillation amplitude (maximum pitch angle in every cycle) and frequency for the cases with $\alpha_g=24.5^{\circ}$ and $\alpha_g=42^{\circ}$ as circles. Time is represented by the darkness of the circle, with darker circles representing the state of the system at later times. We refer to this as the frequency-amplitude trajectory of the system. For the case of $\alpha_g=24.5^{\circ}$ in figure \ref{fig:u40_energy}(a), we see that the system settles into a stationary state in the stable region of the lower $C_E=0$ curve. This is due to the fact that the initial deflection caused by the gust is small enough that the system starts off within the region of negative energy extraction. Therefore the amplitude decays from that initial point to the lower stable equilibrium. On the other hand, the $\alpha_g=42^{\circ}$ gust causes an initial deflection that is just large enough to push the system into the $C_E>0$ region above $A_{\theta} \approx 5^{\circ}$. Due to this, the system continues to extract energy and increase in amplitude of oscillation, until it meets the stable equilibrium boundary at $A_\theta \approx 37^{\circ}$. Thus, the energy map allows us to explain why a $2\times$ difference in $\alpha_g$ leads to a $10\times$ increase in the oscillation amplitude.

\subsection{Tailoring gust response via structural frequency}
\label{sec:f0.16}
The topology of the energy map can also be used to design a system that is robust to specific gust perturbations. As an example, here we demonstrate how a small change in the structural properties of the system described earlier can reduce the stationary state amplitude of the system by a disproportionate amount. By examining the energy map for $X^*_e=0.33$ (see figure \ref{fig:u40_energy}), we see that for $f^*_p \lessapprox 0.16$ the region of negative energy extraction in the low-amplitude region extends to larger $A_{\theta}$ values than at $f^*_p = 0.19$. Hence the energy map suggests that reducing the natural frequency of the system from $f^*_n=0.19$ to $0.16$ should ``trap'' the system in this lower ``island'' of negative energy transfer. In order to verify this, we test a system with natural frequency $f^*_n=0.16$, oscillating about $X^*_e=0.33$, encountering the stronger gust case from the preceding discussion (i.e. $\alpha_g=42^{\circ}$). 

\begin{figure}
  \centerline{\includegraphics[scale=0.9]{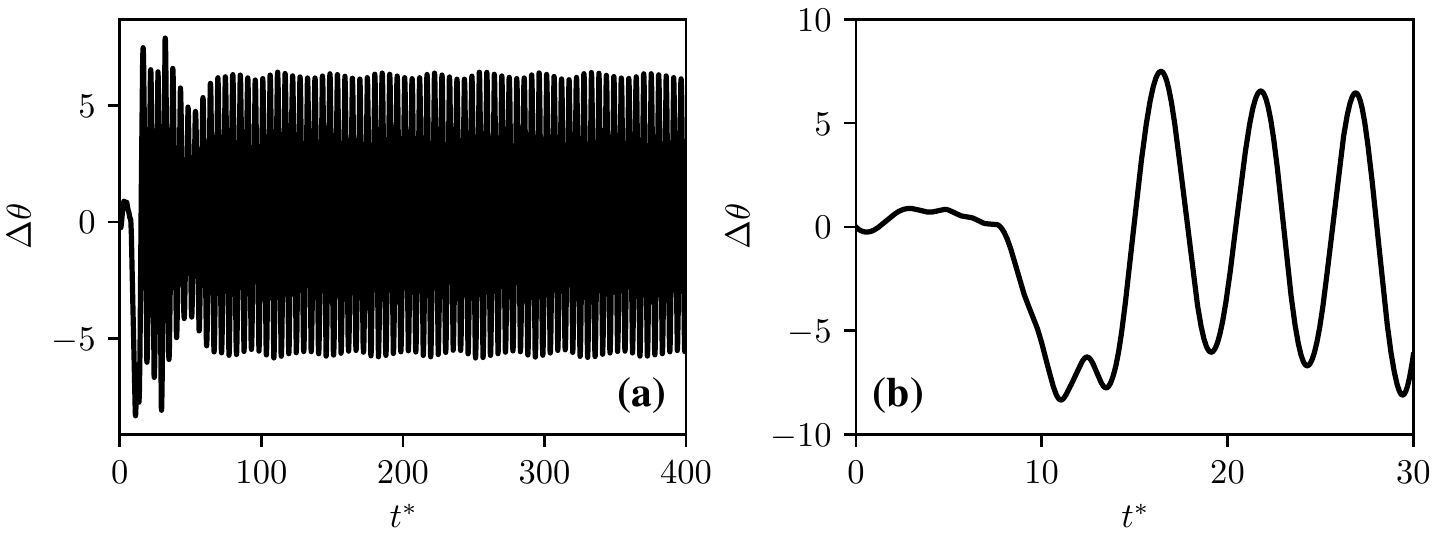}}
  \caption{(a) Time-series of pitch angle for a system with $f^*_n=0.16$, $X^*_e=0.33$, and interacting with a gust of strength $\alpha_g=42^{\circ}$; (b) Zoom-in of the time-series in (a), focusing on the initial deflection induced by the gust.}
\label{fig:u6_pitch}
\end{figure}
In figure \ref{fig:u6_pitch}(a) we plot the time-series of pitch angle for this case and we see that the system settles into stationary state oscillations of $\Delta \theta \approx 6^{\circ}$. As expected, the stationary state amplitude observed is much smaller than the $\Delta \theta \approx 35^{\circ}$ seen for the case with $f^*_n=0.19$. In figure \ref{fig:u6_pitch}(b) we focus on the initial deflection of the system after interaction with the gust. As we did for the previous discussion, we can verify the quasi-steady assumption made in predicting the amplitude of the initial deflection. Since the angle of the gust is $\alpha_g=42^{\circ}$, this corresponds to a steady angle of attack of $\theta_s=57^{\circ}$. The moment time-series for this steady airfoil was shown previously in figure \ref{fig:u40_moment_zoom}(b), and the peak moment for the steady airfoil is $C_M'=0.63$. Since $I^*$ is kept fixed throughout this study, the change in $f^*_n$ corresponds to changing $k^*$. For this case, $k^*=0.0754$ and hence the predicted initial deflection is $\Delta \theta'=8.32^{\circ}$. Comparing this to the actual measured deflection of $\Delta \theta=8.33^{\circ}$, we see that the quasi-steady assumption can again be verified.

\begin{figure}
  \centerline{\includegraphics[scale=1.0]{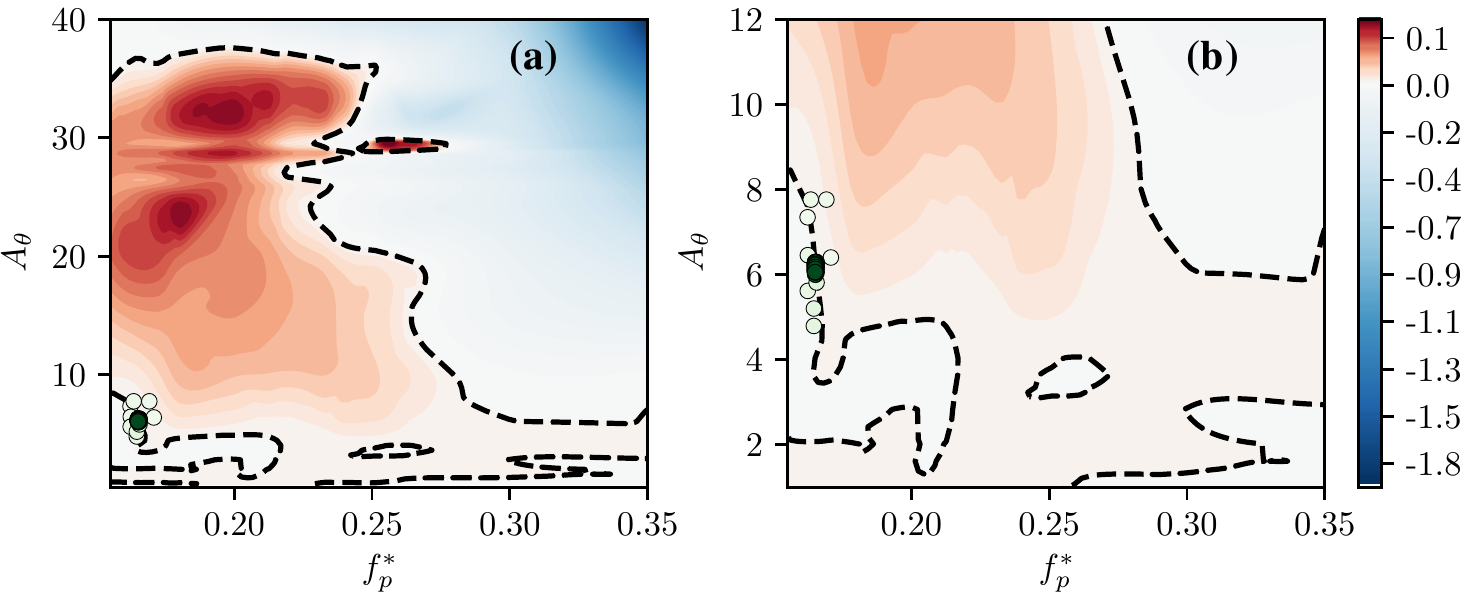}}
  \caption{(a) Frequency-amplitude trajectory plotted on the energy map (for details, see caption under figure \ref{fig:u40_energy}), for a case with $f^*_n=0.16$, $X^*_e=0.33$, and $\alpha_g=42^{\circ}$; (b) Zoom-in of the trajectory plotted in (a), showing the system settling on a region of neutral stability (represented by the nearly-vertical equilibrium curve).}
\label{fig:u6_energy}
\end{figure}
In figure \ref{fig:u6_energy}(a) we plot the frequency-amplitude trajectory for the system on the energy map, and a zoomed-in version of this trajectory is plotted in figure \ref{fig:u6_energy}(b). These figures explain why the system with $f^*_n=0.16$ does not grow to large amplitude oscillations, and shows that the behaviour agrees with what was expected from the change in $f^*_n$. Specifically, the fact that the region of $C_E<0$ extends to larger values of $A_{\theta}$ at this natural frequency causes the amplitude after the initial deflection to stay within this region instead of growing. Interestingly, as seen in figure \ref{fig:u6_energy}(b) the system does not settle into a stable equilibrium in this case but gets trapped on a neutrally stable segment of the $C_E=0$ curve. As mentioned in section \ref{sec:energy_map}, the stability of an equilibrium is given by the sign of $dC_E/dA_{\theta}$ \citep{Morse2009,Menon2019}. A region of $dC_E/dA_{\theta} = 0$ corresponds to neutral stability, where small changes in the oscillation amplitude do not change the sign of $C_E$, and hence do not change the stability of the system. Such a region manifests itself as a vertical contour line of $C_E=0$ on the energy map. Due to this, as seen in the frequency-amplitude trajectory in figure \ref{fig:u6_energy}(b), the system is able to show small variations in oscillation amplitude while remaining at an equilibrium. 

The simulations therefore show that a change of approximately $16\%$ in $f^*_n$ lead to a reduction of about $86\%$ in the flutter amplitude. The simulations also demonstrate how the energy map for a given configuration can inform design changes or control strategies for aeroelastic systems.  In this demonstration, we focused on the reduction in flutter amplitude by decreasing the $f^*_n$, but the same demonstration also works in reverse, i.e. to show how the flutter amplitude could be {\it{increased}} (for instance, in energy harvesting applications) by changing the structural attributes of the system. Note that changes in $f^*_n$ can, in principle be achieved by changing $I^*$, which could be implemented via changes in mass distribution, and/or $k^*$, which could be implemented with the use of smart materials or other such means. This is discussed in section \ref{sec:summary}.

\subsection{Gust response in systems below critical flutter speed}
\label{sec:subcritical}
The energy map also allows us to predict when a gust will destabilise an aeroelastic control surface that is operating below the critical flutter speed. The critical flutter speed, is a key feature in the aero-structural design of aerodynamic control surfaces in air-vehicles \citep{specification1993airplane}. The critical flutter speed is however conventionally determined under the assumption of an infinitesimal perturbation and the analysis employs linear methods. The effect of a finite-size perturbation, which could lead to non-linearity in the response, is therefore not addressed in these analyses. The source of such non-linearity as a result of finite-size perturbations can be either structural or aerodynamic. In the presence of structural non-linearities, such gust-induced subcritical behaviour has been reported by \cite{Dessi2008ASystems} and \cite{Sarkar2008NonlinearVibration}. In the context of energy maps, the possibility of non-linear subcritical responses due to aerodynamic non-linearities is indicated by the presence of multiple stable equilibrium response branches at a given frequency. For a system operating in a stable state below the flutter boundary, a finite perturbation can push the system towards another response branch at a higher $A_{\theta}$ via a subcritical instability. Here we demonstrate this for a case with $f^*_n=0.30$. 

\cite{Menon2019} showed a phenomenological condition for the onset of flutter, and showed that the flutter boundary in terms of frequency is at $f^*_n \approx 0.25$ for the system being studied here. The flutter boundary is often expressed in terms of a critical reduced velocity ($U^* = 1/f^*_n$), which for $f^*_n \approx 0.25$, corresponds to $U^* \approx 4$. The energy map in figure \ref{fig:energy_hinge33} also shows a flutter boundary at $U^* \approx 3.8$ which is indicated approximately by the intersection of a negative energy extraction region with the x-axis at $f^*_p \approx 0.25$. Hence an aeroelastic system with $f^*_n=0.30$ lies on the ``linearly stable'' side of the flutter boundary.
\begin{figure}
  \centerline{\includegraphics[scale=1.0]{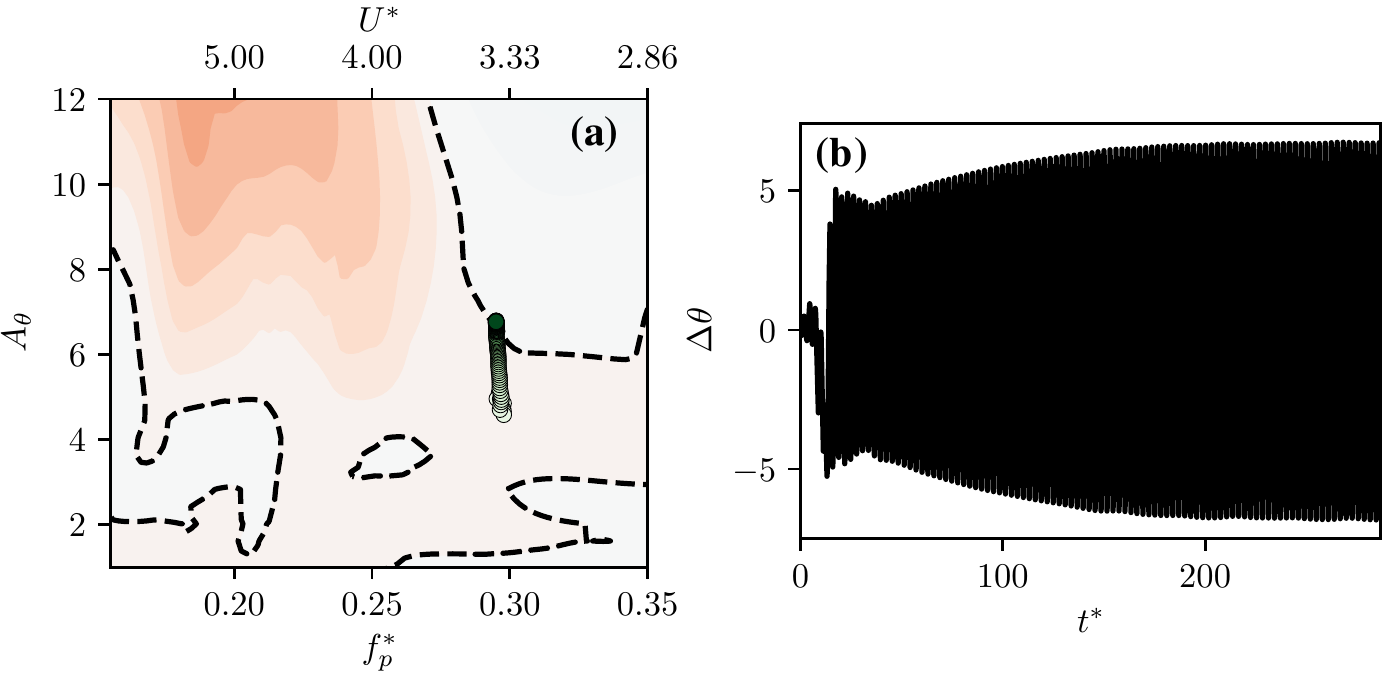}}
  \caption{(a) Trajectory on the frequency-amplitude space for a system with $f^*_n=0.30$, oscillating about $X^*_e=0.33$, and interacting with a gust of strength $\alpha_g=51^{\circ}$. The trajectory is plotted on a zoom-in of the energy map as in figure \ref{fig:u6_energy}(b). For details about the trajectory, see caption under figure \ref{fig:u40_energy}; (b) Time-series of pitch angle for the same case as in (a).}
\label{fig:f30_55_33}
\end{figure}
To demonstrate the subcritical response of the ``linearly stable'' system oscillating with $f^*_n=0.30$ about $X^*_e=0.33$, the airfoil is subjected to an oncoming gust of strength $\alpha_g=51^{\circ}$. Figures \ref{fig:f30_55_33}(a) and \ref{fig:f30_55_33}(b) show the subsequent response of the system on the energy map as well as the time series of pitch deflection. In order to highlight why a subcritical response is expected in this system, figure \ref{fig:f30_55_33}(a) shows the trajectory of the system on the energy map. As we can see, at $f^*_n=0.30$ there exists a region of negative energy extraction at very small amplitude ($A_{\theta} \lessapprox 3^{\circ}$) along with a subcritical stable equilibrium branch at a larger amplitude of $A_{\theta} \approx 7^{\circ}$. The gust perturbation of $\alpha_g=51^{\circ}$ generates an initial pitch perturbation of about $\Delta \theta = 3^{\circ}$ degrees (quasi-steady analysis predicts a pitch excursion of $2.4^{\circ}$) and this relatively small perturbation is sufficient to push the system past the region of stability. The system then finds itself in a state with $C_E > 0$, due to which the amplitude grows until it reaches the stable equilibrium at $A_{\theta} \approx 7^{\circ}$. As can be seen in the time series plot in figure \ref{fig:f30_55_33}(b), the amplitude growth of $\Delta \theta \approx 7^{\circ}$ occurs slower in this system than in the others analyzed thus far, due to the fact that there are multiple equilibrium branches in the vicinity of this frequency, and the energy extraction ($C_E$) is hence small. The fact that long integration times are necessary to compute stationary state responses in flow-induced oscillation systems highlights another advantage of using forced oscillations and energy-based analysis of these systems.

\begin{figure}
  \centerline{\includegraphics[scale=1.0]{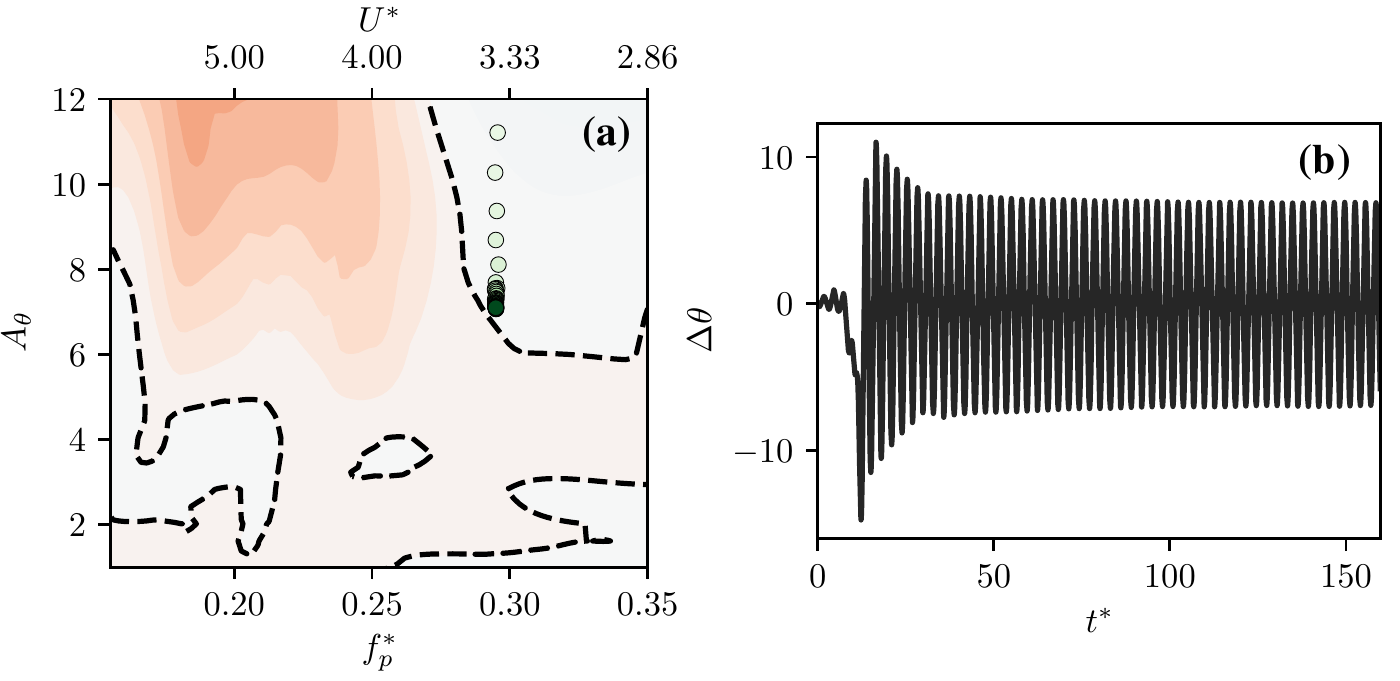}}
  \caption{(a) Frequency-amplitude trajectory plotted on the energy map, for an airfoil with $f^*_n=0.30$, oscillating about $X^*_e=0.33$, and interacting with a gust of strength $\alpha_g=61^{\circ}$. For details about the trajectory, see caption under figure \ref{fig:u40_energy}; (b) Time-series of pitch angle for the same case as in (a).}
\label{fig:u3_33}
\end{figure}
We now demonstrate the robustness of this method in predicting the response to even larger perturbations in this subcritical regime. The system described above, with $f^*_n=0.30$, oscillating about $X^*_e=0.33$, is subjected to a larger gust of strength $\alpha_g=61^{\circ}$, and figure \ref{fig:u3_33} shows the subsequent response of the system. With this gust, figure \ref{fig:u3_33}(a) shows that the system is in fact pushed beyond the upper stable branch, and into a region of negative energy extraction. As expected due to the fact that $C_E < 0$, pitch oscillation subsequently decays until the system attains a limit-cycle state with $\Delta \theta \approx 7{\circ}$. In figure \ref{fig:u3_33}(b) we show the time-series of pitch angle for this case. It must be noted that an amplitude of $\Delta \theta \approx 7^{\circ}$ observed in the two cases discussed on this section might be considered excessive for an air-vehicle. In this context, the current analysis shows that a system operating below the flutter boundary could be destabilized by a gust. Further, it is particularly striking that the angular perturbation required for this destabilization could be as small as $\Delta \theta \approx 3^{\circ}$. On the other hand, for energy harvesting applications, this knowledge could be beneficial in generating large sustained oscillations even in conditions where the system is stationed below the flutter boundary.

\subsection{Tailoring gust response via elastic axis location}
\label{sec:hinge_50}
We now demonstrate a different strategy for controlling the subcritical response of the system. All the cases discussed thus far in this work have been airfoils oscillating about $X^*_e=0.33$. However, \cite{Menon2019} showed that a relatively small change in $X^*_e$ can have a large effect on the topology of the energy map. In particular, the energy map for an airfoil pitching about $X^*_e=0.50$ was studied in that work. Here we aim to leverage this change in topology of the energy map to drive the system to a different stationary state. Note that this does not require any simplifying assumptions, as the energy maps include the full nonlinear behaviour of the system. The natural frequency of the system and the strength of the incoming gust that we analyze in this section are maintained at $f^*_n=0.30$ and $\alpha_g=61^{\circ}$ respectively, to enable direct comparison with the previous case in section \ref{sec:subcritical}.

\begin{figure}
  \centerline{\includegraphics[scale=1.0]{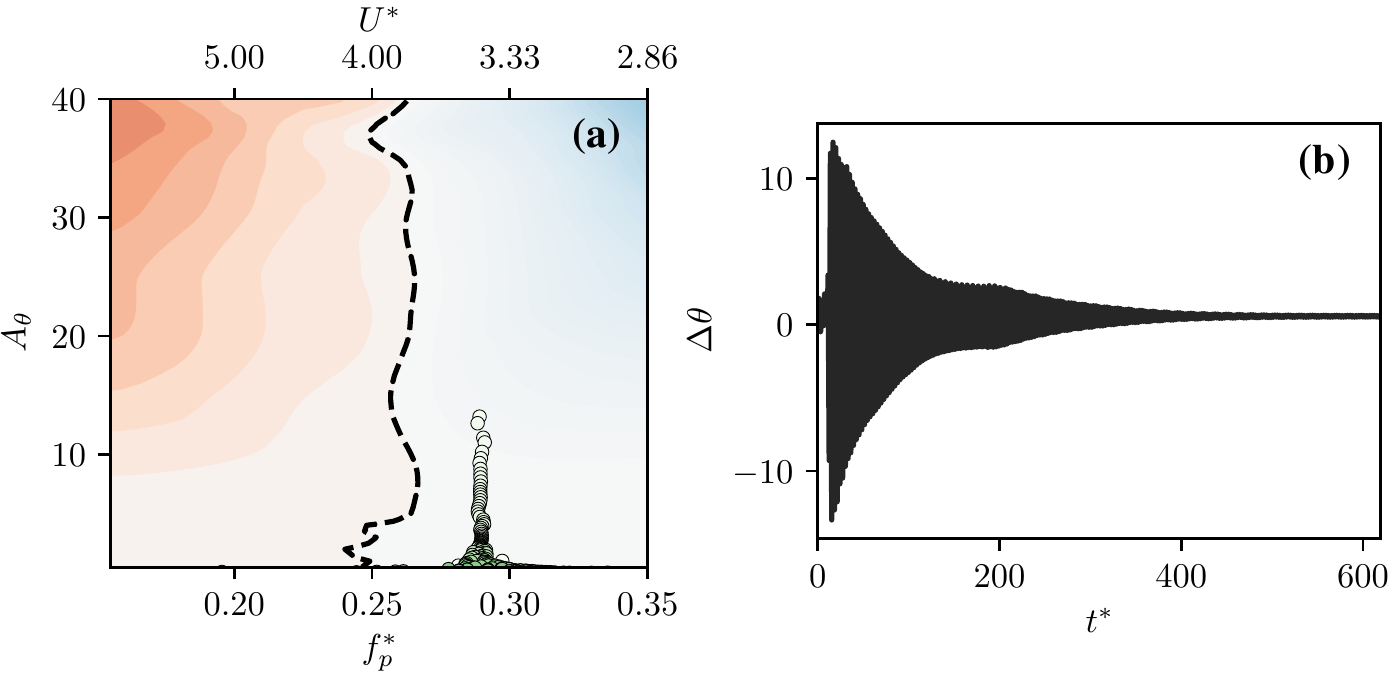}}
  \caption{(a) Trajectory in frequency-amplitude space for a case with $X^*_e=0.50$ (all previous cases discussed have $X^*_e=0.33$), plotted on the energy map corresponding to $X^*_e=0.50$. The natural frequency and gust strength for this case are the same as that in figure \ref{fig:u3_33}, i.e., $f^*_n=0.30$ and $\alpha_g=61^{\circ}$. For details about the trajectory, see caption under figure \ref{fig:u40_energy} (b) Time-series of pitch angle for the same system as in (a).}
\label{fig:u3_50}
\end{figure}
In figure \ref{fig:u3_50}(a) we plot the energy map for an airfoil oscillating about $X^*_e=0.50$, with mean angle of attack $\theta_0=15^{\circ}$ and at $Re=1000$. The plot has the same range of $f^*_p$ and $A_{\theta}$ as the rest of the energy maps discussed in this study. We see that the structure of this map is very different from that for $X^*_e=0.33$. In particular, the flutter boundary is demarcated by a roughly vertical line of $C_E=0$ at $f^*_p \approx 0.25$, as opposed to the complicated structure with multiple equilibrium branches seen for $X^*_e=0.33$ in figure \ref{fig:u3_33}. For the control of subcritical responses this is especially useful, as the simple topology immediately suggests the lack of a subcritical instability at $f^*_n=0.30$. Further, $f^*_n=0.30$ lies in the region of $C_E<0$ for all $A_{\theta}$, which suggests that any perturbations will decay. To verify this claim, the system with $X^*_e=0.50$ and $f^*_n=0.30$ is subjected to a gust of the same strength as the larger of the two cases discussed in the previous section (i.e. section \ref{sec:subcritical} and gust strength $\alpha_g=61^{\circ}$). The trajectory of the system in figure \ref{fig:u3_50}(a) shows that the system oscillations decay due to the negative energy extraction. The time-series of pitch angle shown in figure \ref{fig:u3_50}(b) shows that the initial deflection is of magnitude similar to that of the previous case with $X^*_e=0.33$. However, the final stationary state shows nearly negligible oscillation amplitude of $\Delta \theta \approx 0.5^{\circ}$. 

In summary, the examples in this section as well as the previous one show that a system designed to operate with $X^*_e=0.33$ is susceptible to significant flutter amplitude via a subcritical response to perturbations, even when operating at velocities below the flutter boundary. However, the energy maps suggest that a change in $X^*_e$ forces the system into a strongly stable state, thereby eliminating gust-induced aeroelastic oscillations. Note again that this could work the other way in systems where large oscillations are desirable, such as in energy harvesting. In this case, similar ideas can be used to manipulate the system into the state that is optimal for the application. 

\section{Summary}
\label{sec:summary}

In this study, we have demonstrated a method to accurately predict the response of an aeroelastic control surface to gust-like perturbations. This method is based on energy maps, wherein data from high-fidelity numerical simulations of an equivalent system with prescribed oscillation amplitude and frequency are used to fully characterize the stability characteristics of the system. The topology of an energy map enables us to make fast and accurate predictions of the response of the system to gusts of different strengths and for systems with different structural properties.

As pointed out in section \ref{sec:energy_map}, energy maps have been utilized in previous studies where the assumption of equivalent sinusoidal oscillations in both the forced as well as the flow-induced oscillation has been satisfied. However, here we show that this method can be used even in the presence of transient, non-sinusoidal perturbations induced by gusts. Moreover, a particular merit of this method is that it includes the full effects of aerodynamic non-linearities, which are challenging to model, and often ignored in most simplified aeroelastic models. The current approach therefore extends the applicability of this method beyond the traditional small-amplitude regime, into parameter ranges that are also applicable to energy harvesting applications. Further, this method potentially provides a simple method for flutter testing which obviates the need to perform more expensive flow-induced oscillation tests under a range of different parameters in order to assess the flutter response of a system. By creating an energy map, a potential designer can simultaneously gather results that would be yielded by a traditional flutter response tests (such as trends of limit-cycle oscillation behaviour with respect to velocity), perturbation tests (since the energy map identifies subcritical response branches), as well as possible insight into design conditions. These benefits make the method relevant to a wide range of practical applications ranging from air-vehicles to energy harvesting systems.

We illustrate the efficacy of this energy-map based method in predicting the response of a 2D flow-induced pitching airfoil to transverse gusts. We show that this method allows us to accurately predict the final stationary state of an aeroelastic system, given an initial angle-of-attack deflection that is induced by a perturbation such as a gust. For transverse gusts that are assumed to have length-scales much larger than that of the wing, we find that the initial pitch deflection is governed by the quasi-steady moment corresponding to an effective angle of attack that includes the geometric angle of attack and the gust angle. Hence, given the strength of a transverse gust, this method allows us to estimate the initial deflection and also provides insight into the subsequent stationary-state response of the system. We note that for angle-of-attack perturbations generated by other mechanisms, this energy map-based method is equally applicable and only depends on knowledge of the initial deflection.

We also showed the utility of this method in predicting how changes in gust strength and structural parameters of the oscillator affect the amplitude response of pitching oscillations. Specifically, we showed that we can leverage knowledge of the energy map topology to predict parameter ranges where the system is either more sensitive or more robust to perturbations. This was demonstrated, by showing that the small differences in the initial deflection ($\approx 1^{\circ}$ at $f^*_n=0.19$) can lead to order-of-magnitude differences in the final stationary state oscillation amplitude. Further, we illustrated using examples how the topology of the energy map can inform design modifications to control the stationary state flutter amplitude that results from the interaction with the gust. One such modification involved changing the natural frequency of the system to one where the energy map predicts negative energy extraction for larger oscillation amplitudes than in the previous case. Lastly, we demonstrated the use of the energy map in predicting and controlling the gust response for a system operating below the critical flutter speed. This is also insight provided by the topology of the energy map, due to the presence of subcritical equilibrium curves in certain parameter ranges. In this case, we showed a different way to control the oscillation response by modifying the location of the elastic axis. This enabled us to push the system into a region of negative energy extraction, thus effectively damping the oscillation. Hence the high-degree of aerodynamic non-linearity in the system is seen to manifest itself as a complex energy landscape. We show that a-priori knowledge of the topology of this landscape can be leveraged in multiple ways to gain insight into the response of a system, as well as to inform control strategies.

Having demonstrated the utility of energy maps in studying the perturbation-response of flow-induced oscillators, it is imperative to also assess the limitations of the approach presented in this work. We note that this method relies on a-priori knowledge of the topology of the energy map. Since an energy map for a given shape depends, in principle, on the fluid dynamic conditions (i.e. Reynolds number etc) as well as structural parameters such as natural frequency and elastic axis location, generating a energy map via simulations or experiments might be a daunting task. However, depending on the application and given some a-priori knowledge of the operational envelope of the system, it might be sufficient to generate an a-priori energy map of only a small region of the full parameter space. Another consideration with regards to the use of energy maps in practical situations is the extension of these ideas to systems with multiple degrees-of-freedom. While this is viable, as mentioned in section \ref{sec:energy_map}, it has limitations in terms of the complexity of each mode (and their coupling), as well as the associated computational expense of estimating the energy landscape in higher dimensions. Further, while the development of simple control strategies such a those described in sections \ref{sec:f0.16} and \ref{sec:hinge_50} is feasible in higher dimensions with knowledge of the equilibrium response surfaces of the oscillator, it is likely more complicated. 

In closing, we consider the practical implementation of the ideas presented in this study. The simple strategies presented here to modify the gust response both rely on ``tailoring'' the structural properties of the system (natural frequency and elastic axis in this case). Such on-the-fly structural modifications have been shown to be possible, and beneficial, with the use of smart materials for various aerospace applications \citep{Weisshaar2013MorphingChallenges}, and in particular for flutter suppression \citep{Chattopadhyay1999AeroelasticOptimization,Giurgiutiu2000ReviewControl}. Especially relevant to the strategies explored in this work are various studies exploring the active control of the stiffness distribution in wings \citep{Weisshaar1992AdaptiveIssues,Librescu1997ControlMaterials,Jenett2017DigitalStructures}, which leads to elastic axis and natural frequency modifications. This underscores the practical relevance of the findings discussed in this paper, which we expect will complement aero-structural efforts in the control of flow-induced vibration.

\section*{Acknowledgments}
This work is supported by the Air Force Office of Scientific Research Grant Number FA 9550-16-1-0404, monitored by Dr. Gregg Abate. The development of the flow solver used here has benefited from NSF grant CBET-1511200. This work also benefited from the computational resources at Extreme Science and Engineering Discovery Environment (XSEDE), which is supported by National Science Foundation grant number ACI-1548562, through allocation number TG-CTS100002. Computational resources at the Maryland Advanced Research Computing Center (MARCC) are also acknowledged.

\bibliographystyle{elsarticle-harv} 
\bibliography{references}

\end{document}